# Spin current transport in hybrid Pt / multifunctional magnetoelectric Ga$_{0.6}$Fe$_{1.4}$O$_3$ bilayers

Suvidyakumar Homkar,[†] Elodie Martin,[‡] Benjamin Meunier,[†] Alberto Anadon-Barcelona,[‡] Corinne Bouillet,[†] Jon Gorchon,[‡] Karine Dumesnil,[‡] Christophe Lefèvre,[†] François Roulland,[†] Olivier Copie,[‡] Daniele Preziosi,[†] Sébastien Petit-Watelot,[‡] Juan-Carlos Rojas-Sánchez,*,[‡] and Nathalie Viart*,[†]

[†]*Université de Strasbourg, CNRS, IPCMS, UMR 7504, F-67000 Strasbourg, France*
[‡]*Université de Lorraine, CNRS, IJL, F-54000 Nancy, France*

**ABSTRACT**

The low power manipulation of magnetization is currently a highly sought-after objective in spintronics. Non ferromagnetic large spin-orbit coupling heavy metal (NM) / ferromagnet (FM) heterostructures offer interesting elements of response to this issue, by granting the manipulation of the FM magnetization by the NM spin Hall effect (SHE) generated spin current. Additional functionalities, such as the electric field control of the spin current generation, can be offered using multifunctional ferromagnets. We have studied the spin current transfer processes between Pt and the multifunctional magnetoelectric Ga$_{0.6}$Fe$_{1.4}$O$_3$ (GFO). In particular, *via* angular dependent magnetotransport measurements, we were able to differentiate between magnetic proximity effect (MPE)-induced anisotropic magnetoresistance (AMR) and spin Hall magnetoresistance (SMR). Our analysis shows that SMR is the dominant phenomenon at all temperatures and is the only one to be considered near room temperature, with a magnitude comparable to those observed in Pd/YIG or Pt/YIG heterostructures. These results indicate that magnetoelectric GFO thin films show promises for achieving an electric-field control of the spin current generation in NM/FM oxide-based heterostructures.

**ABSTRACT GRAPHIC**

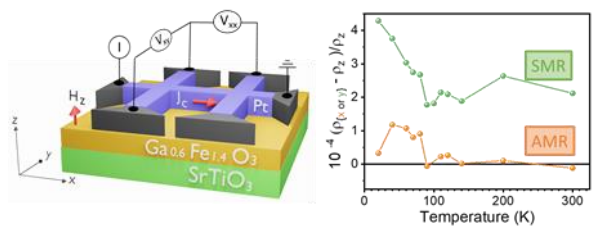

**KEYWORDS:** magnetic oxides thin films, gallium ferrite, platinum, spintronics, spin Hall magnetoresistance, magnetic proximity effects

## 1. INTRODUCTION

Non ferromagnetic heavy metal (NM) / ferromagnet (FM) heterostructures are currently largely investigated to study viable routes to exploit the fascinating interplay between heat, charge and spin transport.[1–3] In particular, recent developments point at a suitable low-power manipulation of magnetization in spintronic devices through spin-orbit torques (SOT) effects, which ground on the intrinsic spin-orbit coupling. Indeed, it has been shown that the spin current generated upon the application of a charge current in the heavy metal nonmagnetic material *via* the spin Hall effect (SHE) can manipulate the magnetic moments in the adjacent ferromagnetic (FM) layer through SOT.[4–10] This has triggered the development of a new Magnetic Random Access Memory (MRAM)-based technology, the spin-orbit torque MRAM (SOT-MRAM)[8], and the possibility to integrate SHE-based spin-valves in standard CMOS-like networks has already been demonstrated via complementary spintronic logic (CSL) design methodology.[11]

The interest for NM/FM heterostructures has recently moved from all-metal systems to systems in which the FM is an insulating oxide (FMI), and display the interesting phenomenon of spin Hall magnetoresistance (SMR).[12–14] The most emblematic NM/FMI system is based on garnet ferrite, with Pt/YIG ($Y_3Fe_5O_{12}$).[3,15–22] However, spinel ferrites have also been considered so far, with systems such as Pt/$NiFe_2O_4$,[3,18] Pt/$Fe_3O_4$[18] and Pt/$CoFe_2O_4$.[23–25] Recently, a study on Pt/$Bi_{0.9}La_{0.1}FeO_3$ bilayers[26] made a step towards the exciting challenge of the electric field control of the spin Hall effect. This, by granting the control of the generation of spin current in engineered NM/FM-based devices, adds extra functionalities to the realization of future spintronics devices.

Here we propose the use of another oxide, the multifunctional magnetoelectric gallium ferrite $Ga_{2-x}Fe_xO_3$ (GFOx, $0.8 \leq x \leq 1.4$),[27,28,29] with a view to use its magnetoelectric character in the future for an electric field control of the spin current generation from NM/FM heterostructures. $Ga_{2-x}Fe_xO_3$ crystallizes in the polar orthorhombic $Pna2_1$ (equivalently $Pc2_1n$) space group (S.G.#33),[30,31] different from the usual perovskite structure adopted by most of the other magnetoelectric compounds, with a = 0.5086(2) nm, b = 0.8765(2) nm, and c = 0.9422(2) nm for x = 1.4.[32] The material is polar with a polarization of ca. 25 $\mu C/cm^2$,[33,34] and ferrimagnetic with a Curie temperature increasing with x and reaching values above room temperature for x=1.3.[28,35,36] Taking into consideration the lattice matching possibilities, thin films of GFOx (00l) can be epitaxially deposited on various substrates such as yttrium stabilized zirconia (001) (YSZ),[37,38] YSZ (111),[38] Pt (111) buffered YSZ (111)[37] or strontium titanate SrTiO$_3$ (111) (STO).[38,39] For symmetry reasons, while the (100) YSZ substrates allow the growth of six in-plane variants, the YSZ (111), Pt (111) buffered YSZ (111) and STO (111) ones will only allow three, the lowest number observed until now. GFO1.4 films were shown to be ferroelectric at room temperature,[39] to have a magnetic Curie temperature of ca. 370 K, and a saturation magnetization of about 100 emu/cm$^3$ at room temperature.[37,40]

## 2. MATERIALS ELABORATION AND CHARACTERIZATION - EXPERIMENTAL DETAILS

Here we show results obtained from Pt/GFO1.4 heterostructures which, for convenience, will be simply indicated as Pt/GFO. These heterostructures were deposited onto SrTiO$_3$ (111) (STO) substrates (Furuuchi Chemical Corporation, Japan, with rms-roughness lower than 0.15 nm), by pulsed laser deposition using a KrF excimer laser ($\lambda$= 248 nm) with a fluence of 4 J/cm². The choice of STO as the substrate was dictated by the will to have the lowest number of in-plane variants, and we have recently demonstrated the possibility to have a layer-by-layer growth of highly crystalline GFO on this substrate.[40] The growth of such atomically flat GFO films is a prerequisite to high quality Pt/GFO interfaces. The GFO layer (ca. 30 nm thick) was deposited first, at 900°C, by ablating a sintered stoichiometric Ga$_{0.6}$Fe$_{1.4}$O$_3$ ceramic target at a repetition rate of 2 Hz of in a 0.1 mbar O$_2$ pressure, using an already optimized procedure described elsewhere.[40] The overall composition of the film has been assessed by energy dispersive X-ray spectroscopy (EDX) coupled to a scanning electron microscopy technique (JEOL 6700 F). The analysis was performed at 5 keV, ensuring a large surface sensitivity of the EDX signal. The Pt deposition was performed at a repetition rate of 10 Hz, under the system base pressure of 2.10$^{-8}$ mbar, and at room temperature to avoid any interdiffusion between the metal and oxide layers. The Pt thickness chosen here for generating spin currents was of 5 nm. This thickness has been evidenced by other works as optimum for SHE-induced spin currents from its associated spin diffusion length ($\lambda_{sd}$), spin memory loss (SML) at the interface[41] and spin Hall torque efficiency per applied electric field unit.[42] The surface of the STO//GFO/Pt heterostructure, observed by atomic force microscopy (AFM) with a Bruker ICON microscope operated in tapping mode, has a low rms roughness of 0.3 nm (Figure 1a). The roughness of a GFO layer of similar thickness (32 nm) deposited alone on a STO (111) substrate, without any

Pt layer on top, had already been previously characterized, and is also of about 0.3 nm.[40] Observations of the heterostructure cross section by transmission electron microscope (TEM) (JEOL 2100 F) allowed confirming the low roughness of the Pt/GFO interface. A composition profile was measured across the interface by energy dispersive X-ray spectroscopy (EDS) and showed that interdiffusion is limited in this system and restricted to a less than 2 nm wide zone (Figure 1b). X-ray diffraction was performed on the heterostructure with a Rigaku Smart Lab diffractometer equipped with a rotating anode (9 kW) and a monochromated copper radiation (1.54056 Å). Both GFO oxide and Pt metal layers of the heterostructure are well crystallised (Figure 1c). GFO is oriented along its [001] direction and Pt, along its [111] direction, on the STO (111). A layer-by-layer growth together with a smooth Pt/GFO interface are demonstrated by clear Laue oscillations for both GFO and Pt around their 004 and 111 reflections, respectively. Reflectivity measurements allowed us to determine that the precise thicknesses of the GFO and Pt layers are 36 and 5 nm, respectively. The magnetic properties of the heterostructure were studied with a superconducting quantum interference device vibrating sample magnetometer (SQUID VSM MPMS 3, Quantum Design). Temperature dependent measurements of the magnetization performed in field cooled and zero field cooled modes (not shown here) indicate a Curie temperature of 364 K. The sample indeed still shows a ferromagnetic behaviour at room temperature with a saturation magnetization of 100 emu/cm$^3$, as expected for this Fe/Ga ratio.[37] A highly anisotropic behaviour is evidenced from the in-plane and out-of-plane magnetization loops measurements (Figure 1d). The magnetization lies preferentially in-plane, in perfect agreement with the fact that the [100] and [001] crystallographic axes are, respectively, the easy and hard magnetic directions for GFO,[28] and are observed from X-ray characterizations to lie, respectively, in- and out-of-plane.

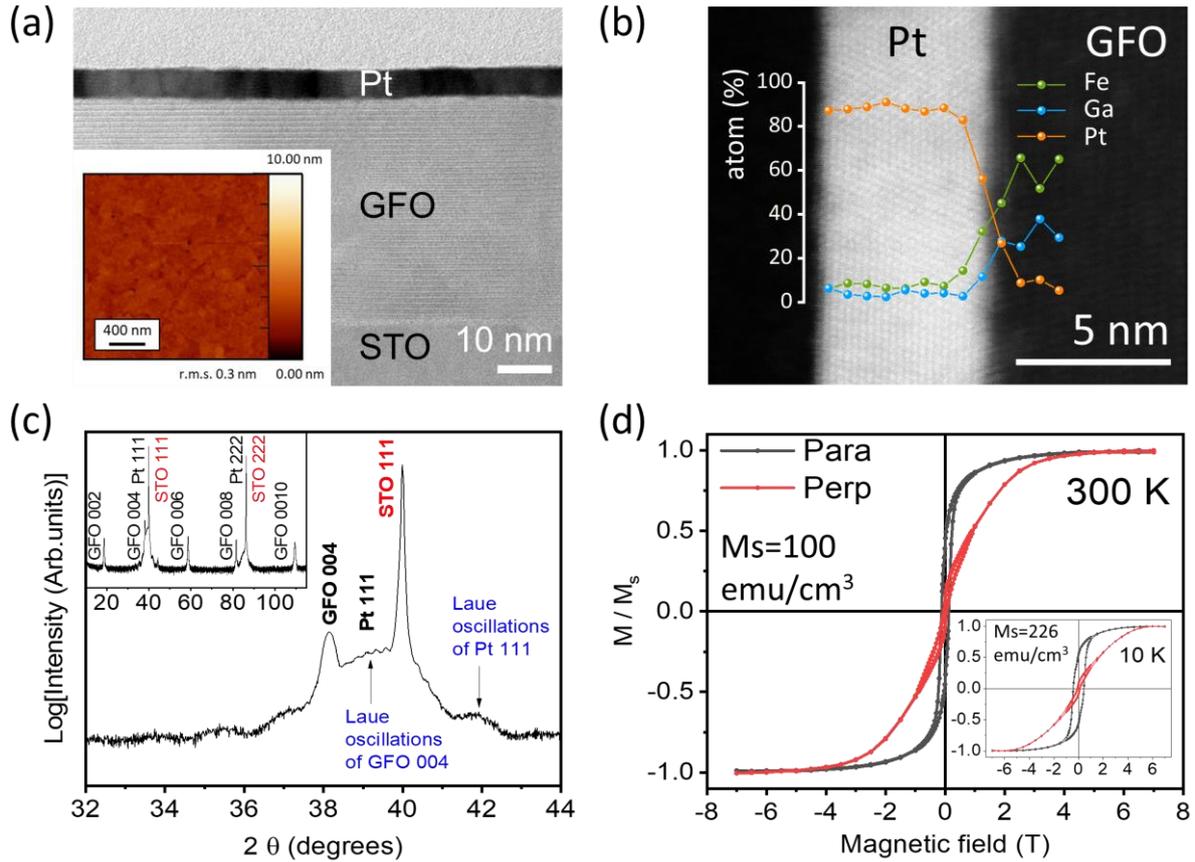

Figure 1. (a) TEM cross view of the STO//GFO(36 nm)/Pt(5 nm) heterostructure with (insert) the AFM image of its top surface indicating a rms roughness of 0.3 nm, (b) EDX mapping at the Pt/GFO interface, (c) X-ray diffractogram of the heterostructure in the θ-2θ mode, with a focus on the GFO 004 peak showing the Laue oscillations observed for both the GFO 004 and Pt layers 111 reflections, (d) Magnetization hysteresis loops of the heterostructure measured in both parallel and perpendicular modes at 300 K and (insert) 10 K.

## 3. MAGNETOTRANSPORT STUDY – RESULTS AND DISCUSSION

The Pt layer was patterned, using standard optical lithography, into double Hall bars to minimize the electrical resistance associated to the metallic contacts, with a longitudinal length L=38 µm, a width w=10 µm, and a thickness t=5 nm (Figure 2). The current is injected in the Pt bar along the STO[0-11] (ΓM) direction, that is, perpendicularly to the [100] easy magnetization direction of GFO. The plane of the films will be referred to as $xy$, and the normal to the film will be named $z$. $H_z$ ($H_x$, $H_y$) refers to the magnetic field applied along the $z$ ($x$, $y$) direction. The electric current density $J_c$ is applied along $x$, and the longitudinal (transversal) resistivity $\rho_{xx}$ ($\rho_{xy}$) is calculated from the measured voltage $V_{xx}$ ($V_{xy}$).

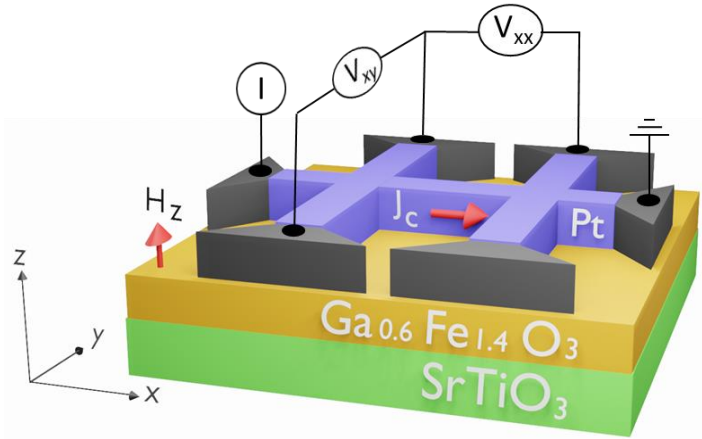

Figure 2. Schematics of the lithographed double Hall bars on STO//GFO(36 nm)/Pt(5 nm) heterostructures with indications of length (L), width (w) and thickness (t).

Magnetotransport in heavy metal (NM) / insulating ferromagnet (FMI) heterostructures may involve contributions from two main origins: (1) the magnetic proximity effect (MPE), and (2) the spin Hall related effects with the spin Hall magnetoresistance (SMR). Figure 3 offers a schematic representation of the microscopic mechanisms behind each contribution and for two different geometries of measurement, *i.e.* longitudinal and transverse.

Going to details, MPE implies the existence of some induced interfacial magnetism in the non-magnetic Pt layer. As a result, anisotropic magnetoresistance (AMR) and anomalous Hall effects[43] (AHE), characteristic of metallic FM, can play a role.

Instead, SMR contribution does not necessitate the existence of any MPE as already put forward by Nakayama *et al.*[14] SMR effects arise as a combination of the spin Hall effect (SHE) and the inverse spin Hall effect (ISHE). When a charge current flows longitudinally in the nonmagnetic NM, a spin current is produced along the film normal direction by SHE. The spin polarization $\vec{\sigma}$ of this spin current is perpendicular to both charge and spin current densities, $\vec{J_e}$ and $\vec{J_s}$, respectively, in agreement with $\vec{J_s} = \theta_{SH}(\vec{\sigma} \times \vec{J_e})$, where $\theta_{SH}$ is the spin Hall angle.[44,45] The spin current can either be reflected or absorbed by the adjacent FM layer depending on whether $\vec{\sigma}$ is parallel or perpendicular to the magnetization direction of the FM layer, respectively.[18] The reflected spin current will produce an additional charge current through the inverse spin Hall effect (ISHE) which will lead to a decrease of the longitudinal resistivity. The resistivity of the NM layer will therefore strongly depend upon the orientation of the FM magnetization.

If one neglects contributions from other phenomena such as the topological Hall effect[46] (THE) or unidirectional magnetoresistance[47,48] (UMR), very unlikely in this collinear-spins system, the longitudinal ($\rho_{xx}$) and transverse ($\rho_{xy}$) resistivities of the studied heterostructure as a function of an external magnetic field can be described by the following equations:[13,18]

$$\rho_{xx} = \rho_0 + \Delta\rho_{MPE\ AMR} \cdot m_x^2 + \Delta\rho_{||SMR} \cdot m_y^2 \qquad (1)$$

$$\rho_{xy} = (\Delta\rho_{MPE\ PHE} - \Delta\rho_{||SMR}) \cdot m_x \cdot m_y + (\Delta\rho_{MPE\ AHE} + \Delta\rho_{\perp\ SMR}) \cdot m_z \qquad (2)$$

where $m_x$, $m_y$, and $m_z$ are the magnetization unit vector components in the *x*, *y*, and *z* directions, respectively, $\rho_0$ is the resistivity of platinum in the absence of a magnetic field, $\Delta\rho_{MPE\ AMR}$ is the anisotropic magneto-resistivity due to the MPE induced anisotropic magnetoresistance (MPE AMR), and $\Delta\rho_{MPE\ PHE\ (AHE)}$ is the planar (anomalous) Hall resistivity due to the MPE induced planar (anomalous) Hall effect in the NM. The AMR and AHE contributions, which depend on the orientation of the MPE-induced magnetic Pt, arise due to extrinsic spin-flip scattering mechanism and/or intrinsic mechanism depending on the band structure of Pt. In addition, one has to consider the SMR related phenomena. $\Delta\rho_{||SMR}$ represents the SMR effect and $\Delta\rho_{\perp\ SMR}$ is a Hall-effect-type resistivity that can be relevant in systems where the imaginary part of the spin mixing conductance is important, but it is usually smaller than $\Delta\rho_{||SMR}$.[18]

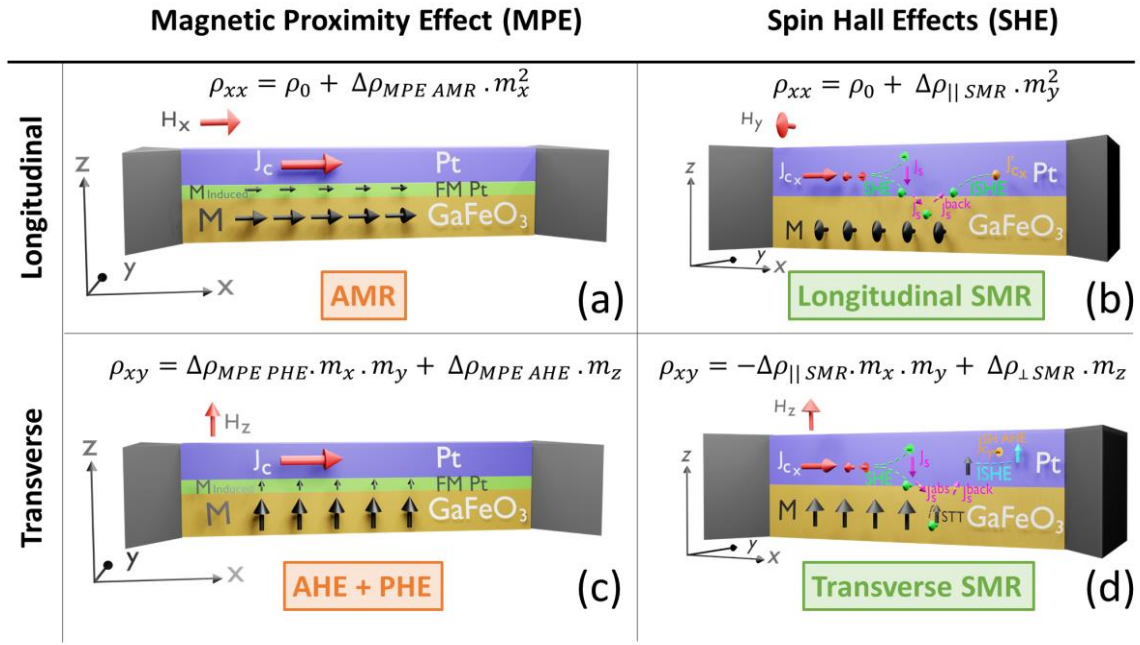

Figure 3. Schematics of the various physical phenomena which have to be considered for magnetotransport in a Pt/GFO heterostructure, for both longitudinal, (a) and (b), and transverse, (c) and (d), modes involving the magnetic proximity effects (MPE) with anisotropic magnetoresistance (AMR), anomalous Hall effect (AHE) and planar Hall effect (PHE), and the spin Hall induced effects (SHE).

**Longitudinal magnetoresistance**

The longitudinal magnetoresistance (MR) is defined as $\frac{\rho_{xx\,(H)} - \rho_{xx\,(H=7T)}}{\rho_{xx\,(H=7T)}} = \frac{\Delta\rho_{xx}}{\rho_{xx}}(H)$. The field dependence of the MR (for a field applied along the z direction) is plotted in Figure 4a for various temperatures between 20 and 300 K. These curves are a clear evidence of the presence of non-zero longitudinal magnetoresistance (MR). The MR decreases with increasing temperature, goes to zero at approximately 120 K, after what it keeps a negligible value (Figure 4b). This behavior of MR cannot be attributed to weak localization (WL) and weak anti-localization (WAL) mechanisms. If one cannot completely exclude the existence of a 2DEG at the STO/GFO interface, it will however not be possible to observe any of its transport properties in the Pt layer, where the electrical contacts are made, because of the insulating character of the GFO layer between the STO and Pt layers. Moreover, the temperatures at which WL and WAL are at play are usually much lower than 120 K, since such quantum effects require that the coherence of the wave functions is kept. Possible origins behind the temperature evolution of the MR will be discussed later. The MR curves presented in Figure 4a do not saturate even at magnetic fields beyond the magnetic saturation of the heterostructure observed by SQUID (Figure 1d). This is probably caused by some independent moments at the interface.[49] It can be related to the highly anisotropic nature of magnetism in the GFO films, which induces high anisotropy for the induced magnetic Pt as well, through a magnetic proximity effect, and will inhibit a saturation point even at high temperatures. The measured Pt resistivity value of 23 µΩ.cm at room temperature is in good agreement with values reported for Pt thin films deposited either by sputtering or molecular beam epitaxy.[50,51] Its linear decrease with temperature is also in agreement with previous studies.[52]

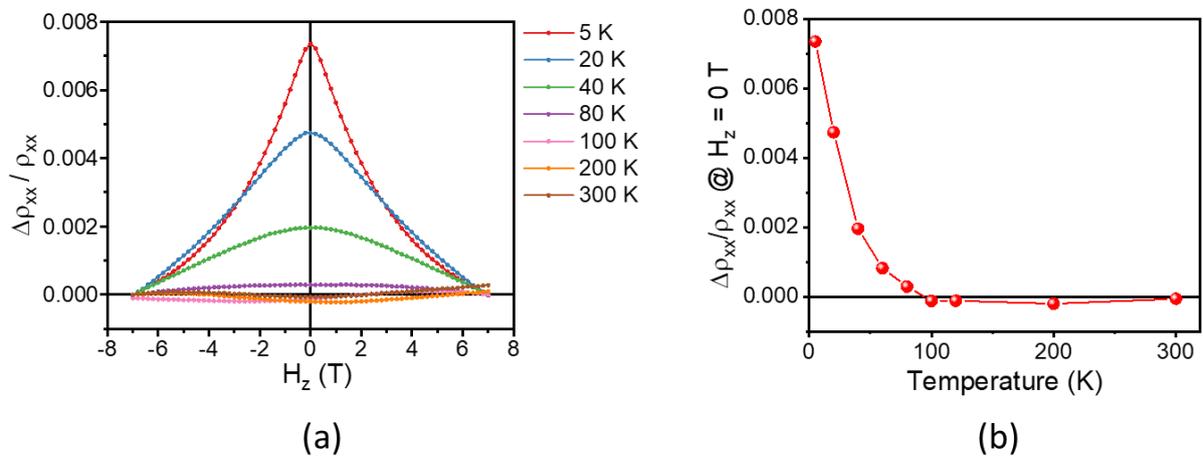

Figure 4. Longitudinal magnetoresistance (as defined in the main text). (a) Measurements at various temperatures, and (b) Temperature dependence of the longitudinal magnetoresistance estimated for $H_z$= 0 T.

**Transverse magnetoresistance**

The magnetic field dependence of the transverse Hall resistivity $\rho_{xy}$ measured at various temperatures between 20 and 300 K (the field is applied along the z direction), is presented in Figure 5a, after correction from both the ordinary magneto-resistance (OMR) contribution due to the variation of the temperature and the ordinary Hall resistance (OHR) due to the Lorentz force applied onto the carriers. For all temperatures, this corrected transverse resistivity $\rho_{xy-corr}$ behaves as an odd function (opposed signs for opposed Hz fields). In the light of Equation (2), this means that it results from the contribution of the second term, $(\Delta\rho_{MPE\ AHE} + \Delta\rho_{\perp\ SMR})$. The first one, associated to in-plane projections of magnetization, is not expected to reverse with reversing Hz fields.

One can observe a sign reversal of the $\rho_{xy-corr}$ signal with temperature on Figure 5a. A more precise estimation of the temperature at which this sign reversal operates can be done by plotting the temperature dependence of the values measured for $\rho_{xy-corr}$ at 7 T (Figure 5b). The inversion temperature is of about 120 K, which is the same as the one at which the longitudinal resistivity $\rho_{xx}$ goes to zero (Figure 4b). A sign inversion of the transverse resistivity has already been observed for Pt/YIG systems and assigned to AHE-induced by MPE in the Pt.[15,19,53] Similar temperature variations of the AHE were also observed in the absence of any FM layer, in ion-gated platinum thin films and analogously attributed to some induced ferromagnetic ordering on the Pt surface.[16,54] The explanation given by Zhou *et al.*[53] for such a sign inversion is that for paramagnetic Pt, both the density of states (DOS) and curvature near the Fermi surface importantly change with temperature. They indeed observed, for Pt/YIG heterostructures, a sign inversion of the ordinary Hall coefficient $R_0$ of Pt with temperature, indicating a change of the carriers type. The raw resistivity curves we measured, before OMR and OHR corrections, are shown in SI, Figure S1. The ordinary Hall coefficient of Pt, $R_0$, determined as the slope of the linear part of the measurements at high magnetic fields, is always negative and does not vary significantly with temperature. It allows determining a free electron density in Pt of 48 .$10^{28}$ /m$^3$ , in good agreement with values expected for a metal[55] and already reported for other Pt thin films.[15,16] The absence of any sign inversion of $R_0$ in our case could originate from the fact that we have used a thicker Pt layer (5 nm) than the one used by Zhou *et al.* (1 nm).[53] Shimizu *et al.*,[16] who have used 3.5 nm thick Pt layers, also observe a constant negative $R_0$. The modification of the DOS and curvature near the Fermi surface, induced by a MPE, is indeed expected to happen only at the interface between the NM and FM materials, and could be masked in our case, for which the bulk signature predominates. We cannot therefore be fully conclusive on the incidence of MPE-induced AHE on the temperature behaviour of transverse resistivity.

In order to go further in the understanding of our system, we have sought to disentangle the MPE-based and SMR contributions in the longitudinal measurements.

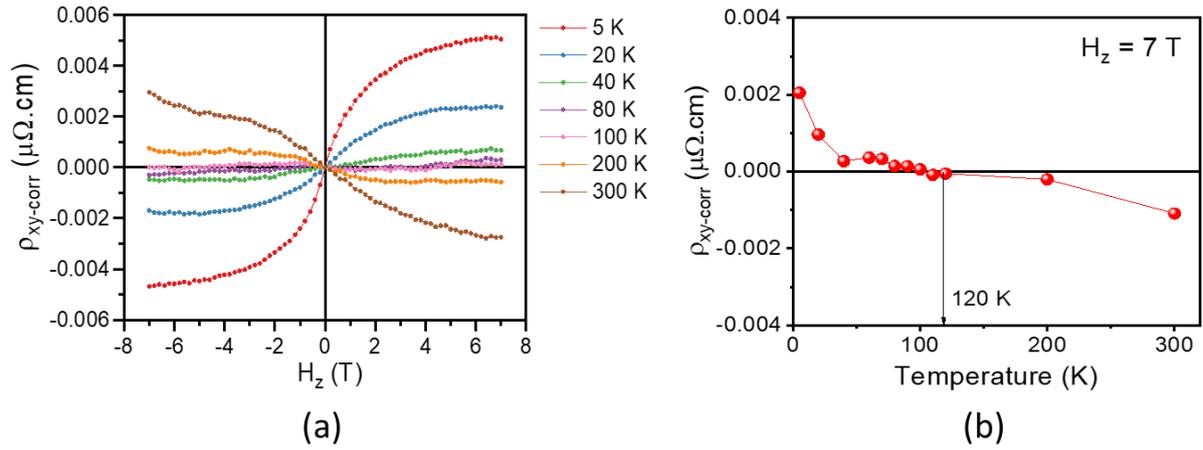

Figure 5. Transverse Hall resistivity measurements with (a) Transverse resistivity corrected from both the ordinary magnetoresistance and the linear contribution of the ordinary Hall effect (for a selection of temperatures), (b) Temperature dependence of the transverse resistivity at $H_z$ = 7 T.

**Angular dependence of the longitudinal resistivity**

One possible way to distinguish between the MPE-based AMR and SMR contributions is through the insertion of a nonmagnetic metal such as Cu or of an antiferromagnetic oxide such as NiO, which might eliminate the possibility of MPE effects, leaving only SMR effects to play a role. However, this introduces additional interfacial issues and important modifications of the SMR, depending upon the interlayer thickness, have been reported in both Pt/Cu/Co/Pt[56] and Pt/NiO/YIG[57] heterostructures.

An alternative way to separate AMR and SMR contributions is by performing angle-dependent longitudinal measurements with the magnetic field in either the *xz* or the *yz* planes, while the current density $J_c$ and the measured resistivity $\rho_{xx}$ are in the *x* direction[49] as schematized in Figure 6a. This can be understood from Equation 1: a rotation of the magnetic field within the *yz (xz)* plane will only have an effect on the SMR (AMR) and no effect on the AMR (SMR) which only depends on $M_x$ ($M_y$). As depicted in Figure 3, while the AMR results from the fact that M is parallel to the current direction, the SMR originates from the fact that M is parallel to the spins of the electrons of the SHE-generated spin current, which prevents the spins to be absorbed by the FM, and causes them to be reflected into Pt.[14]

We measured $\rho_{xx}$ at H = 7 T while rotating the sample in the *yz (xz)* plane to study the $\beta$ ($\alpha$) angle dependence of $\rho_{xx}$, at various temperatures. The orientation of the magnetic field is described through the α (within the *xz* plane) and β (within the *yz* plane) angles. For the experiments, the α and β angles were limited to a 90° rotation, and conventionally, the z direction was chosen as the 90° angle. Both $\frac{\rho(\beta)-\rho_z}{\rho_z}$, and $\frac{\rho(\alpha)-\rho_z}{\rho_z}$ calculated quantities are shown in Figure 6b for various temperatures. The β-dependent measurements show that the resistivity value decreases when going from $H_z$ to $H_y$ at all temperatures, with a 180° periodic oscillation. On the other side, the α measurements also show that the resistivity value decreases when going from $H_z$ to $H_x$ but with a smaller change in resistivity, and no unambiguous periodic oscillation. By fitting the β measurements with $\cos^2\beta$ (SI, Figure S2), the $\frac{\rho_y-\rho_z}{\rho_z}$ SMR values can be extracted. The α measurements could not unambiguously be fitted with $\cos^2\alpha$, and the $\frac{\rho_x-\rho_z}{\rho_z}$ MPE AMR values are extracted from the $\frac{\rho(\alpha)-\rho_z}{\rho_z}$ measurements at α = 0°. This procedure is comforted by the fact that the values of $\frac{\rho_y-\rho_z}{\rho_z}$ (SMR), extracted from the β measurements, have the same values as $\frac{\rho(\beta)-\rho_z}{\rho_z}$ for β = 0°. Hence, assuming that the α measurements are periodic as well, we extracted the $\frac{\rho_x-\rho_z}{\rho_z}$ (AMR) values from $\frac{\rho(\alpha)-\rho_z}{\rho_z}$ measurements at α = 0°.

The SMR and AMR values obtained as explained are plotted in Figure 6c. Both contributions globally decrease with increasing temperatures. SMR shows a minimum at about 120 K, which is also the temperature at which AMR goes to zero. For all temperatures, the SMR contribution dominates over the AMR one, and it is the only one present above 120 K. The predominance of the SMR mechanism over the entire temperature range has also been observed for Pt/YIG and Pd/YIG samples. The SMR value measured for the Pt/GFO heterostructures is about $2 \cdot 10^{-4}$ at 300 K and $4.5 \cdot 10^{-4}$ at 20 K. These values are similar to what is observed in Pd/YIG heterostructures, and only slightly smaller than the ones observed in Pt/YIG ($4 \cdot 10^{-4}$ at 300 K and $6 \cdot 10^{-4}$ at 20 K).[49] If the AMR contribution we observe is very similar to the one reported in other works[19,49,58,59]: same positive sign, similar amplitude of ca. $10^{-4}$, and a decrease with increasing temperature which leaves it practically insignificant after about 100 K, we however highlight some differences stemming from the temperature dependence of the SMR contribution. Studies performed under a high magnetic field, such as the present one of Pt/GFO performed at 7 T or the study of Pt/YIG under 100 kOe,[49] show V-shaped SMR curves, with first a decrease and then an increase with the increasing temperature, the position of the minimum varying between 20 and 120 K from one system to another. The measurements performed in lower magnetic fields (10 kOe)[53,59] globally show an inversed tendency, with first an increase and then a decrease, with a maximum at about 100 K.

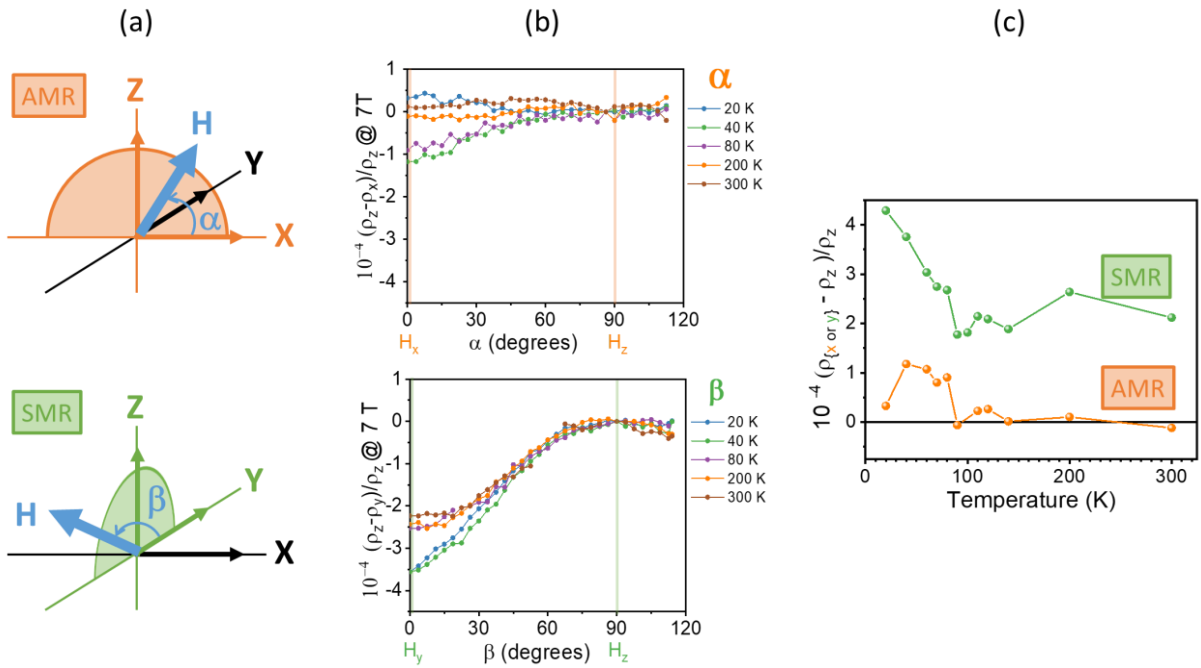

Figure 6. Identification of the SMR and AMR contributions to the longitudinal effects, (a) Geometries of the measurements, (b) Angular measurements at various temperatures (selection of temperatures) in both geometries, (c) Temperature dependence of SMR and AMR as deduced from the $\frac{\rho(\beta)-\rho_z}{\rho_z}$ and $\frac{\rho(\alpha)-\rho_z}{\rho_z}$ values measured at zero degree angle for β and α, respectively.

The magnetic state of the magnetic material is probably to be considered to explain such differences. Recently, a theoretical study has shown that the orbital hybridization of the magnetic material plays a role in the magnetoresistance, most probably in relation with spin-orbit coupling.[60] The minimum in SMR we observe here could thus be put in perspective with the modification of the spin-orbit coupling observed in bare GFO thin films near 120 K *via* our XMCD study.[61]

In fact, since both SMR and MPE-induced magnetoresistive contributions stem from the interaction between the charge current flowing in the Pt layer and the magnetic properties of the FM, the modification of the spin-orbit coupling at 120 K for GFO could contribute to the observed temperature variations of both longitudinal and transverse magnetoresistance measurements in the 90-140 K temperature range (Figures 4b and 5b).

## 4. CONCLUSION

The multifunctional magnetoelectric GFO oxide has successfully been introduced in FMI/NM (with Pt as the NM) heterostructures of high crystalline quality. The Pt/GFO interface is sharp and this makes the heterostructure suitable for spin currents transparency. The interactions between the spin Hall current from Pt and the GFO magnetic orientation have been evidenced by magnetotransport measurements. SMR has been shown to be the dominant phenomenon at all temperatures, and it is the only one to be considered near room temperature, with a magnitude comparable to those observed in the classically studied Pd/YIG or Pt/YIG heterostructures. This study therefore validates the use of GFO as a multifunctional magnetoelectric material in NM/FMI heterostructures with a view to control their spin current generation by an electric field.

## SUPPORTING INFORMATION

Transverse Hall resistivity measurements, Fitting of the longitudinal angular dependent measurements

## AUTHOR INFORMATION

**Corresponding Authors**

*E-mail: juan-carlos.rojas-sanchez@univ-lorraine.fr (J.-C.R.-S.)

*E-mail: viart@unistra.fr (N.V.)

## ACKNOWLEDGEMENTS

This work was funded by the French National Research Agency (ANR) through the ANR-18-CECE24-0008-01 'ANR MISSION' and, within the Interdisciplinary Thematic Institute QMat, as part of the ITI 2021 2028 program of the University of Strasbourg, CNRS and Inserm, it was supported by IdEx Unistra (ANR 10 IDEX 0002), and by SFRI STRAT'US project (ANR 20 SFRI 0012) and ANR-11-LABX-0058_NIE and ANR-17-EURE-0024 under the framework of the

French Investments for the Future Program. The authors wish to thank D. Troadec (IEMN, Lille, France) and A.-M. Blanchenet (UMET, Lille, France) for the preparation of the TEM FIB lamellae, as well as the XRD, MEB-CRO, and TEM platforms of the IPCMS. We acknowledge partial support from the French PIA project "Lorraine Université d'Excellence", reference ANR-15IDEX-04-LUE. Devices in the present study were patterned at MiNaLor clean-room platform which is partially supported by FEDER and Grand Est Region through the RaNGE project.